# The density of states of graphene underneath a metal electrode and its correlation with the contact resistivity

Ryota Ifuku, *Kosuke Nagashio, Tomonori Nishimura, and Akira Toriumi
Department of Materials Engineering, The University of Tokyo, Tokyo, 113-8656 JAPAN
*Email: nagashio@material.t.u-tokyo.ac.jp

The density of states (*DOS*) of graphene underneath a metal is estimated through a quantum capacitance measurement of the metal/graphene/$SiO_2$/$n^+$-Si contact structure fabricated by a resist-free metal deposition process. Graphene underneath Au maintains a linear *DOS* - energy relationship except near the Dirac point, whereas the *DOS* of graphene underneath Ni is broken and largely enhanced around the Dirac point, resulting in only a slight modulation of the Fermi energy. Moreover, the *DOS* of graphene in the contact structure is correlated with the contact resistivity measured using devices fabricated by the resist-free process.

To make the best use of the extremely high carrier mobility of graphene in electric devices, contact resistivity ($\rho_c$) should be seriously addressed[1,2] because it is necessary to be lowered by several orders of magnitude from the present status.[1] Generally, the current injection from metal to graphene is proportional to the transmission probability and the density of states (*DOS*) in the metal and graphene.[3,4] In terms of the transmission probability, an intrinsic problem exists for momentum matching because the Dirac cone at the K point of the Brillouin zone boundary is generally lager than the Fermi wavenumber for typical metals.[5] The number of empty states in graphene as the final state is much smaller than the number of occupied states in the metal. This *DOS* bottleneck is considered to be the limiting factor in the total performance of graphene devices.[6,7] This limitation results from the inapplicability of conventional doping techniques such as ion implantation[8] because the thermodynamically stable C-C bonding prevents the substitutional doping of B or N for C.[9] The *DOS* of graphene in the contact structure, i.e., the metal/graphene/$SiO_2$ structure, could be different from the ideal linear relation because the electrical properties of graphene with monoatomic thickness are easily modulated by the environment. Therefore, determining the *DOS* of graphene in the contact structure provides information on the strength of the metal/graphene/$SiO_2$ interaction, which is key to setting guidelines to further reduce $\rho_c$.

Here, in the case of graphene grown on metals, the modulation of the linear dispersion is reported to depend strongly on the metal element used, according to both theoretical calculations[10,11] and experimental results.[12,13] The modulation in the electron dispersion relation occurs on a chemisorption group (e.g., Ni, Co, and Pd) and not on an adsorption group (e.g., Au, Ag, and Pt).[10] Based on these reports, we selected Ni as a contact electrode to measure $\rho_c$[1] because the *DOS* of graphene in the contact structure is expected to increase due to the strong $\pi$-$d$ coupling.[14] However, in the graphene field effect transistors (FETs) fabricated using the conventional resist process, it has been shown that graphene underneath the Ni electrodes roughly maintains linear dispersion based on the transport measurements.[15]

This discrepancy could be caused by the resist residue in the device fabrication process. Many researchers have reported that the resist residue remains on graphene.[16-18] The resist residue is a serious concern in light of the fact that activated carbon, whose hydrophobic surface attracts organic materials, is composed of graphene.[19] Indeed, in the case of the direct deposition of Ni on graphene, Ni(111) grew epitaxially, whereas Ni was amorphous or had a tiny crystalline structure on graphene after the conventional resist process.[15]

In this report, the resist-free graphene/metal interaction is studied for two typical metals, Ni and Au, using the *DOS* - energy relationship determined by the quantum capacitance ($C_Q$) measurement of the metal/graphene/$SiO_2$/$n^+$-Si stack system, as shown in Fig. 1(a). The $C_Q$ of graphene extracted from capacitance measurements provides direct information on the *DOS* of graphene because it is regarded as the energy cost of inducing carriers in graphene and is directly related to the *DOS* according to $C_Q = e^2 DOS$.[20,21] Although many other techniques, such as photoemission spectroscopy[12,13] and scanning tunneling spectroscopy[22], have been used to extract the *DOS* of graphene, these techniques cannot be applied to the contact

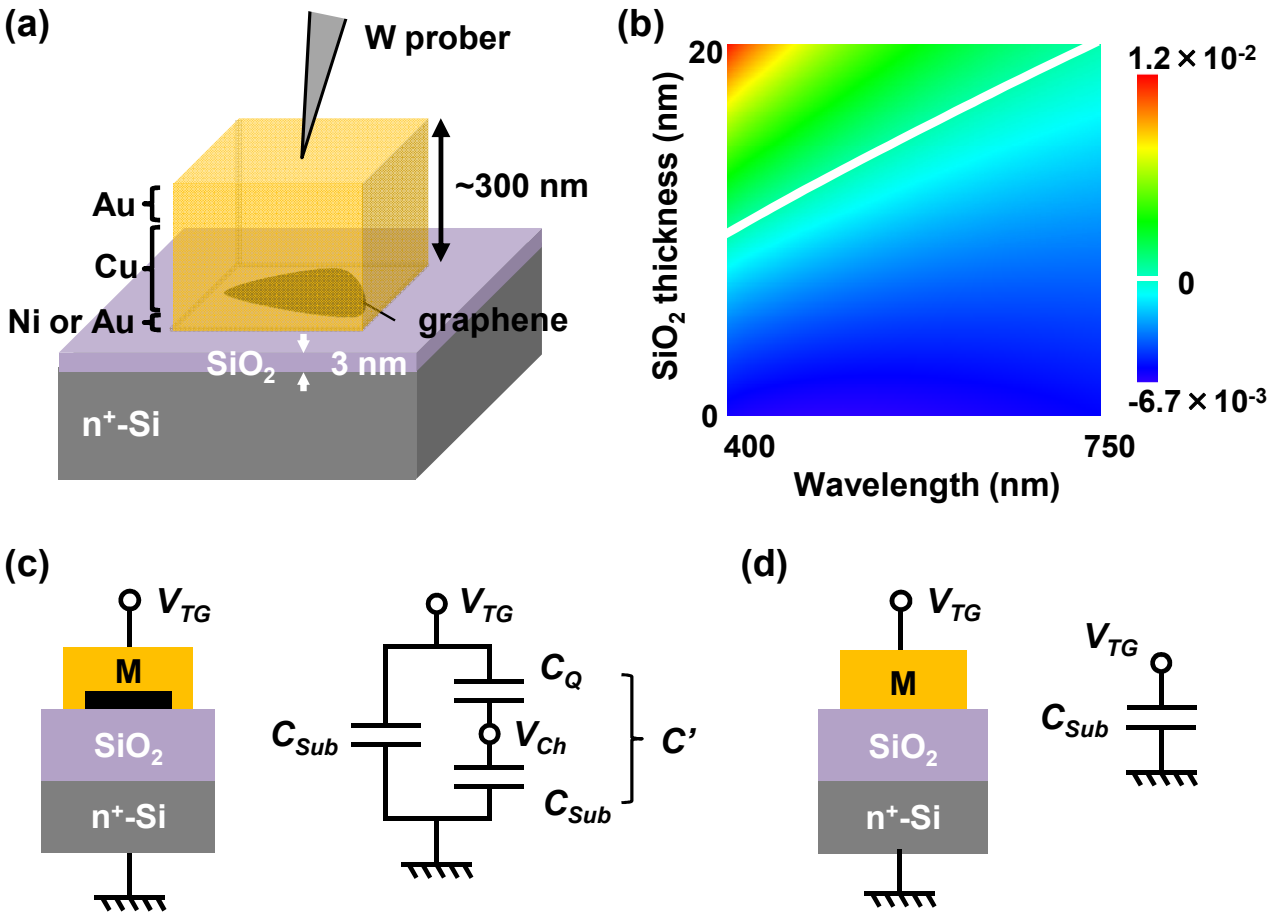

**FIG 1** (a) Schematic of the metal/graphene/$SiO_2$/$n^+$-Si device. (b) Color plot of the optical contrast for graphene on the thin $SiO_2$/Si substrate calculated as functions of wavelength and $d_{SiO2}$. The white line indicates the zero contrast. (c) Schematic of the capacitor with graphene (side view) and its equivalent circuit. (d) Schematic of the capacitor without graphene (side view) and its equivalent circuit.

structure due to the lack of an accessible graphene surface. Moreover, in this report, $\rho_c$ is determined in the graphene FETs fabricated using the resist-free metal deposition technique. Finally, the correlation between the *DOS* of graphene in the contact structure and $\rho_c$ is discussed.

There is a critical requirement for the $SiO_2$ thickness ($d_{SiO2}$) in the metal/graphene/$SiO_2$/$n^+$-Si capacitor in Fig. 1(a) in order to extract $C_Q$.[23] Because $C_Q$ is introduced in series with the geometrical capacitance ($C_{ox}$) in the equivalent circuit ($1/C = 1/C_{ox} + 1/C_Q$), $C_{ox}$ should be at least comparable to $C_Q$. Therefore, $C_Q$ has so far only been extracted in top gate devices with a thin high-*k* oxide layer (<10 nm).[21,24-26] In this study, the thickness of the $SiO_2$ layer in metal/graphene/$SiO_2$/$n^+$-Si structure is adjusted to be thinner than 5 nm. Although a thickness of 90 or 280 nm for $SiO_2$ is generally selected to efficiently visualize graphene, we have noticed that the optical contrast exists even for thin $SiO_2$ layers. Figure 1(b) shows the color plot of the contrast as functions of wavelength and $d_{SiO2}$ calculated using the Fresnel equation assuming a trilayer model of graphene, $SiO_2$, and Si.[27] Although the optical contrast (< 0.01) for graphene on thin $SiO_2$ is generally weaker than that on 90-nm $SiO_2$ (~0.1), a thinner $d_{SiO2}$ provides better visibility for thicknesses less than ~10 nm. Meanwhile, the gate leakage current caused by direct tunneling increases exponentially with decreasing oxide thickness, especially for thicknesses less than ~2 nm for the $SiO_2$/Si substrate. Therefore, a $SiO_2$ layer with ~ 3-nm thickness is used in this study.

Graphene was mechanically exfoliated from Kish graphite onto ~ 3-nm $SiO_2$/$n^+$-Si substrates (0.01 Ωcm). Graphene was detected after the contrast was optimized by adjusting the RGB values of the image, which revealed that the graphene contrast was indeed brighter than that of $SiO_2$, as suggested by the negative contrast in Fig. 1(b). The contours of graphene were clearly observed using an optical microscope in dark-field mode.[23] The monolayer graphene was confirmed by Raman spectroscopy. A motorized position-alignment system was used for the resist-free metal deposition, as shown in Fig. 2(a). The pre-patterned poly(methyl methacrylate) (PMMA) stencil mask, supported by the Si substrate with a 200-µm square window (Fig. 2(b)), was adjusted to the position of graphene on the $SiO_2$/$n^+$-Si substrate under the optical microscope.[23] Next, this PMMA stencil mask was fixed to the $SiO_2$/$n^+$-Si substrate with tape. The topgate contact metal was thermally evaporated under vacuum at a background pressure of $10^{-4}$ Pa. A three-layered metal stack (composed of a Au or Ni; contact layer ~ 20 nm thick, a Cu protection layer ~ 200 nm thick, a Au; probing layer ~60 nm thick) was used to prevent mechanical damage to graphene by the direct probing to the metal contact, as shown in Figs. 1(a) and 1(c). Moreover, it is difficult to precisely deposit the metal along the contours of graphene. The devices without graphene in Fig. 1(d) were also fabricated using the same metal deposition steps for comparison, as observed in Fig. 2(b). Alternatively, the PMMA resist (495 PMMA A11, MicroChem Corp.) was intentionally spin-coated onto graphene and removed by acetone to clarify the effect of the resist residue. After the adjustment of the mask position, the metals were deposited onto graphene. The capacitance measurements were performed at room temperature in ambient air at a frequency of 1 MHz to suppress the leakage current and to reduce the slow time-constant response. Although the resistance was concerned to be high near the Dirac point (DP) due to the very small *DOS*, it was confirmed by a simple *RC* consideration that 1 MHz was low enough because the carrier density near the DP in graphene on $SiO_2$ is large enough (~$10^{11}$ $cm^{-2}$) due to the charged impurities.[21,28]

Moreover, backgate graphene FETs with a device structure appropriate for the transfer length method (TLM) were fabricated by the resist-free metal deposition technique to determine $\rho_c$. The long as-transferred graphene (~50 µm) was selected for use in this study, as shown in Fig. 2(c). A SiN membrane mask patterned by a focused-ion-beam apparatus, as shown in Fig. 2(c), was used for the *I-V* measurements instead of the PMMA mask that was used for the *C-V* measurements because the hardness of SiN enables a more complex pattern. The pad patterns were also included in this mask. Because the graphene channel region was exposed to air after the metal deposition unlike in the case of the process used with the *C-V* devices, all of the devices were annealed in vacuum at 300 °C for 1

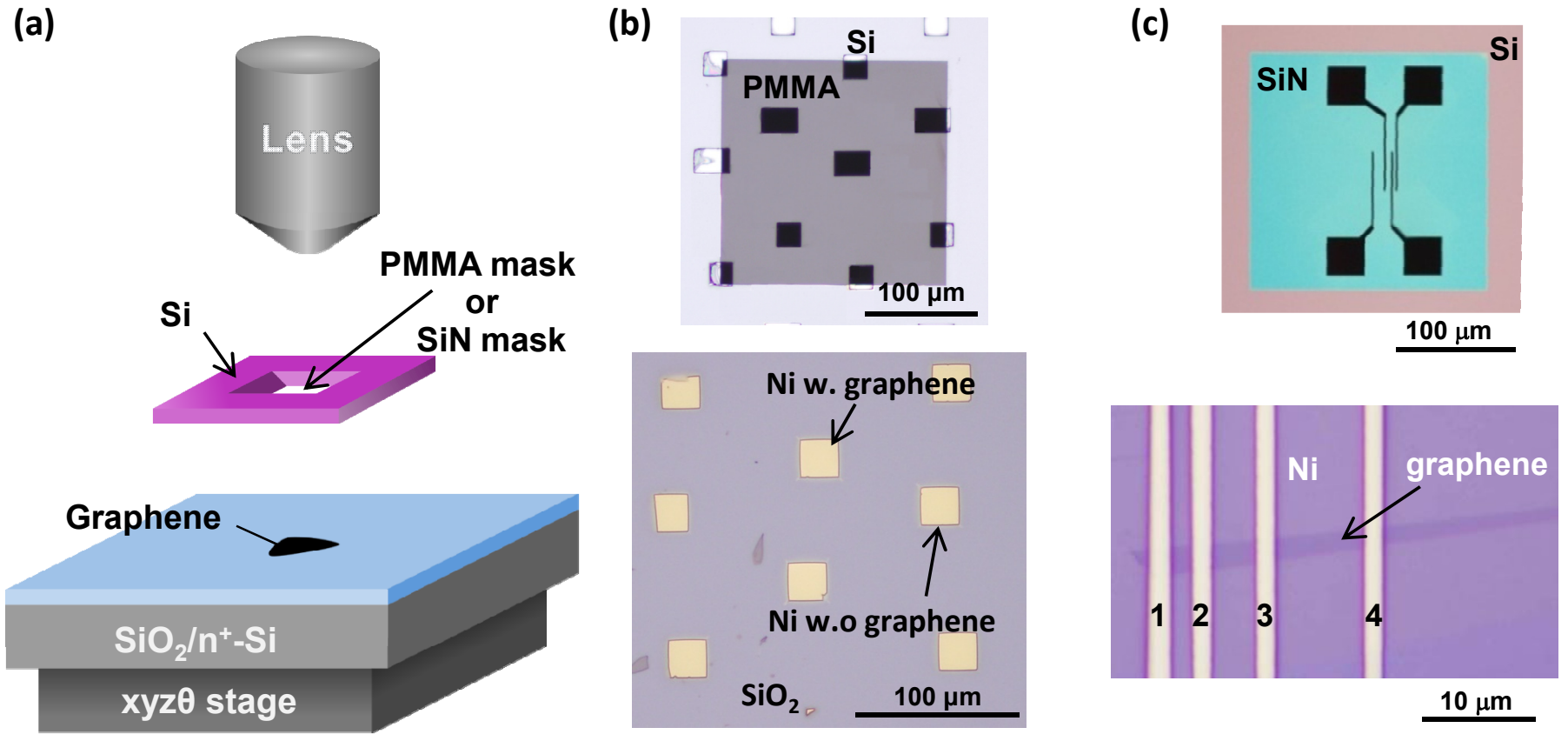

**FIG 2** (a) Schematic of the position adjustment system for the PMMA mask and graphene. The position of the PMMA mask pattern can be adjusted relative to that of graphene by the xyzθ stage driven by the stepping motor. (b) (top) The PMMA stencil mask supported by the Si substrate with a 200-µm square window. (bottom) Optical image of the *C-V* device with and without graphene. (c) (top) The SiN stencil mask supported by the Si substrate with a 200-µm square window. The TLM pattern was fabricated by the FIB instrument. (bottom) Optical image of the *I-V* device with the TLM structure fabricated by the resist-free metal deposition technique. The electrodes are always numbered sequentially from 1, which represents the shortest channel.

hour and the electrical measurements were performed in vacuum at room temperature.

First, the effect of the resist residue on the metal/graphene interaction is addressed. Figures 3(a) and 3(b) show the capacitance as a function of the topgate voltage ($V_{TG}$) for the resist-free and resist-processed Ni devices, respectively. The blue data were obtained from the device with graphene in Fig. 1(c), while the red data were obtained from the device without graphene in Fig. 1(d). The broad capacitance change observed in all of the data as a function of $V_{TG}$ is a result of the formation of the depletion layer in the $n^+$-Si substrate. In addition, the dip attributed to the $C_Q$ of graphene is clearly observed near $V_{TG}$ = 0 V in the resist-processed Ni device in Fig. 3(b), whereas there is no dip in the resist-free Ni device in Fig. 3(a).

The $C_Q$ of graphene in contact with Ni was extracted. The equivalent circuits of two types of measured capacitors are modeled in Figs. 1(c) and 1(d), where $C_{Sub}$ is the series capacitance of $SiO_2$ and $n^+$-Si and $V_{Ch}$ is the channel voltage. To extract $C_Q$, the $C_{sub}$ measured for the device without graphene (red data) was first subtracted from the total capacitance of the device with graphene (blue data) after careful determination of the ratio of the metal area on $SiO_2$ to the area on graphene. Thus, the series capacitance ($1/C' = 1/C_Q + 1/C_{Sub}$) was obtained. Note that the parasitic capacitance is not shown in the equivalent circuits in Fig. 1 because the parasitic capacitance can be removed by this subtraction, which is the main advantage of this method. Finally, $C_{Sub}$ was again subtracted from $C'$ to extract $C_Q$. In this $C_Q$ analysis, the keys to obtain accurate data are the following two techniques. One is to increase the area of graphene compared with that of the metal; therefore, when the monolayer graphene was found on ~3 nm-$SiO_2$, the PMMA mask pattern was selected from the pre-fabricated PMMA masks with different pattern sizes to fit the graphene's size. The other is to precisely measure the graphene area, which was achieved using the clear contours in dark-field mode.

Figure 3(d) shows the $C_Q$ extracted for the Ni devices as a function of the Fermi energy ($E_F$). The right vertical axis indicates the *DOS* calculated by $C_Q = e^2 DOS$. $E_F$ is indeed the charging energy and is expressed as $E_F = eV_{Ch}$. When $V_{TG}'$ is defined as $V_{TG}' = V_{TG} - V_{DP}$, $V_{Ch}$ can be expressed as $V_{Ch} = V_{TG}' - \int_0^{V_{TG}'} C'/C_{Sub}\, dV_{TG}'$.[26] $V_{DP}$ is the DP voltage that is determined by the minimum value in the dip. Thus, the experimentally estimated $C_Q$ can be compared with the theoretical $C_Q$ ($=2e^2 E_F/\pi(v_F\hbar)^2$), where $v_F$ is the Fermi velocity ($1\times10^8$ cm/s) and $\hbar$ is Planck's constant. In addition, the $C_Q$ estimated for the $Y_2O_3$ top-gate device in which graphene is sandwiched between the oxides is also added as a reference in this figure.[21] The resist-processed Ni device exhibits a slightly broken ambipolar $E_F$ modulation, while the *DOS* for the resist-free Ni device increases near the DP and the range of $E_F$ modulation is very limited. It should be noted that the gate voltage range is the same for both devices (±1.5 V). This result suggests that the utilization of the organic resist considerably weakens the graphene/metal interaction and also explains the ambipolar behavior in the previous *I-V* measurement of graphene underneath the Ni electrode.[15]

The intrinsic difference between the Ni and Au in the metal/graphene interaction is examined using devices fabricated by the resist-free process. Figure 3(c) shows the capacitance as a function of $V_{TG}$ for the resist-free Au device. Compared with the dip for the Ni device in Fig. 3(b), the dip observed near $V_{TG}$ = ~0 V for the Au electrode is much larger. Therefore, the $C_Q$ extracted for the resist-free Au device in Fig. 3(d) exhibits a large ambipolar behavior similar to the $Y_2O_3$ topgate device, suggesting that Au has little influence on the *DOS* in graphene. Using the resist-free metal deposition technique, it is evident that the interaction of graphene with Ni is much stronger than that with Au. The schematic in Fig. 3(e) summarizes the metal/graphene interaction.

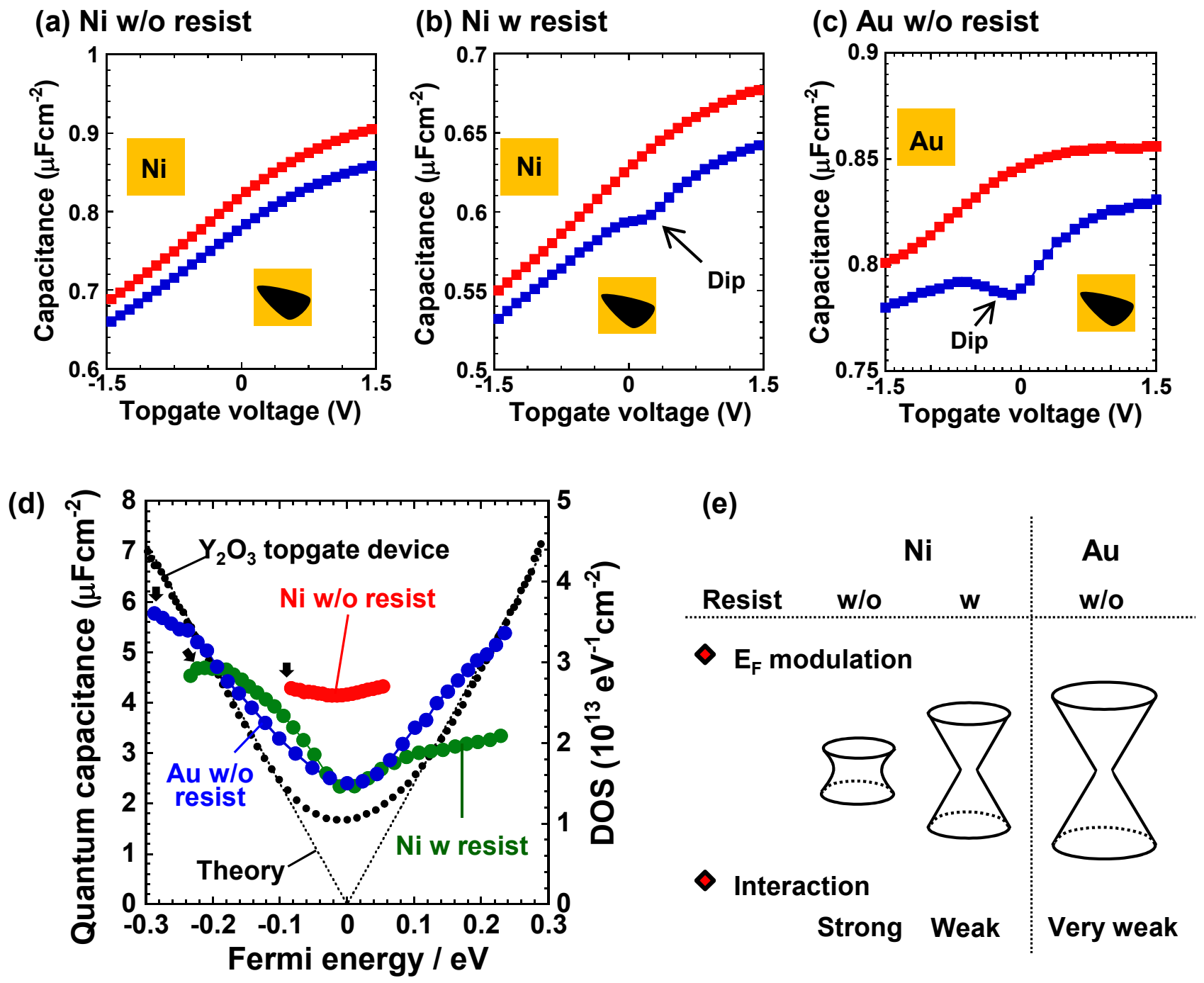

**FIG 3** Total capacitance as a function of $V_{TG}$ for the (a) resist-free Ni, (b) resist-processed Ni and (c) resist-free Au devices. The blue data is obtained from the device with graphene in Fig. 1(c), while the red data is obtained from the device without graphene in Fig. 1(d). (d) The experimentally extracted $C_Q$ of graphene with the theoretical line. Solid circles with green, red, and blue colors represent the resist-processed Ni, resist-free Ni, and resist-free Au devices, respectively. The black circle represents the $C_Q$ of graphene obtained from the $Y_2O_3$ topgate device as a reference.[21] The arrows indicate the highest *DOS* for the three devices. (e) Summary of the metal/graphene interaction suggested from the $C_Q$ measurements.

Next, the electrical transport is discussed for the TLM devices fabricated by the resist-free metal deposition technique. Figures 4(a), 4(b) and 4(c) show the two probe resistances as a function of the backgate voltage ($V_{BG}$) for different channel distances in the resist-free Ni, the resist-processed Ni, and the resist-free Au devices, respectively. The pair of numbers indicates the electrode number shown in Fig. 2(c). When the effect of the resist residue on the metal/graphene interaction is addressed by comparing both Ni electrode cases, there are two clear differences: (i) the DP shift due to the overlapping of the charge-transfer region (*e*-doping) and (ii) the asymmetry of the resistance due to *p-n* junction formation.[29] For (i), Figure 4(d) shows $V_{DP}$ as a function of the channel distance for both Ni electrode devices. These data are fitted by the exponential curve $V_{DP} = -A\exp(-L_{CT}/B)$, where $L_{CT}$ is the charge transfer length and $A$ and $B$ are constants. The value extrapolated to zero $V_{DP}$ indicates the value of $2\times L_{CT}$. The $L_{CT}$ for the resist-processed Ni device was ~0.6 μm, roughly consistent with the previous report in the resist-processed device.[30] The $L_{CT}$ of ~1.3 μm for the resist-free Ni device was much longer. For (ii), based on the $C_Q$ measurement, the $E_F$ of graphene underneath the Ni electrode can be modulated in the resist-processed case, while the modulation is very limited in the resist-free case. The *p-n* junction formed near the contact always exists for $V_{BG} < 0$ V for the resist-free Ni device, which results in the additional resistance. On the other hand, the *p-n* junction can be released for large negative $V_{BG}$ for the resist-processed Ni device, as shown in Figure A3 of the supplementary material.[23] Therefore, the origins of the large DP shift and the large asymmetry of the resistance are the combination of the large charge transfer and the limited $E_F$ modulation.

For the estimation of $\rho_c$, TLM was used for the resist-free Au device.[23] However, TLM is not valid for the Ni device because the channel resistance is not equivalent for all channels with different lengths due to the large DP shift. Therefore, four probe measurements were performed for the Ni devices. Figure 4(e) summarizes the values of $\rho_c$ obtained at $V_{BG} = \pm 30$ V in the present study. Although a considerable improvement in $\rho_c$ was expected by employing the resist-free process for Ni as a result of the increase in the *DOS* due to the *π-d* coupling, no significant improvement was observed. In contrast, $\rho_c$ for the resist-free Au device was the lowest of all the previously reported values.[2]

Finally, the correlation between the *DOS* of graphene and $\rho_c$ is addressed. The *DOS* and $\rho_c$ at exactly the same $E_F$ cannot be compared in this study due to the different $SiO_2$ thickness for the *C-V* and *I-V* devices. Because the relation between $E_F$ and the carrier density ($n$) is expressed as $E_F = \hbar v_F\sqrt{\pi n}$, the $E_F$ at $V_{BG} = 30$ V for the device with the 90-nm thick $SiO_2$ layer is ~0.3 eV, suggesting that the gate voltage ranges for the *C-V* and *I-V* devices are comparable. Therefore, the highest *DOS* at a high $E_F$ was selected as shown by the arrows in Fig. 3(d), whereas the $\rho_c$ obtained at $V_{BG} = \pm 30$ V was used. Figure 4(f) shows the linear relationship between $\rho_c$ and the inverse of the *DOS*. The reduction of the distance between graphene and the metal due to the lack of resist residue generally leads to an increase in the transmission probability. Because the contribution to $\rho_c$ from the *DOS* of graphene and the distance is linked, the dominant factor that reduces $\rho_c$ is not clear yet. As observed from Fig. 4(f), however, it is true that the low $\rho_c$ for the resist-free Au device is due to both the reduction of the graphene/Au distance and the largest *DOS* achieved by the large $E_F$ modulation.

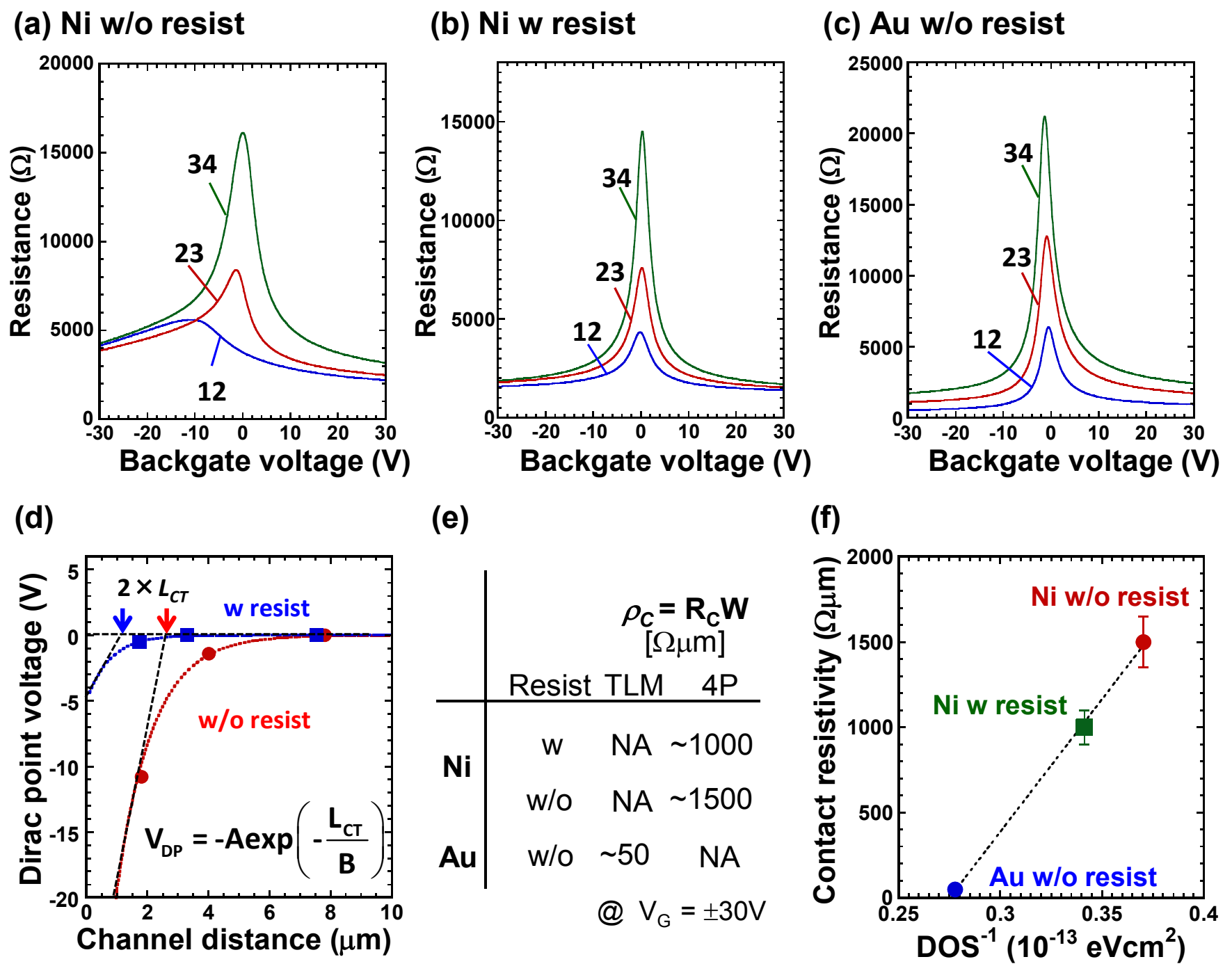

**FIG 4** Two probe resistances as a function of $V_{BG}$ for different channel lengths for the (a) resist-free Ni, (b) resist-processed Ni and (c) resist-free Au devices. The pair of numbers indicates the electrode number shown in Fig. 2(c). (d) $V_{DP}$ as a function of the channel distance for Ni electrode devices. $L_{CT}$ is the charge transfer length. (e) Summary of the $\rho_c$ values determined in the present study. (f) Relationship between $\rho_c$ and the inverse of the *DOS*, where the largest *DOS*, shown by the arrows in Fig. 3(d), was selected.

It should be emphasized that the low $\rho_c$ achieved in the resist-free Au device is not suitable for the topgate device because the *DOS* of graphene underneath the metal electrode cannot be modulated by the topgate. Therefore, we initially hoped for a large increase in the *DOS* based on the strong *π-d* coupling for Ni because this can be used even for the topgate device structure. However, a large increase was not observed, suggesting that the selection of the contact metal might not help in reducing $\rho_c$. The use of graphitic contact formation could be one solution for reducing $\rho_c$ because the *DOS* of graphite is much larger than that of graphene and because better contact properties have been reported for carbon nanotubes.[31]

In this work, using the resist-free metal deposition technique, the *DOS* of graphene in the contact structure was estimated from the $C_Q$ measurement. The utilization of organic resist in the device fabrication process was confirmed to weaken the graphene/metal interaction. For the resist-free metal/graphene contacts, the *DOS* of graphene is maintained for Au, while it is largely modulated for Ni. Although the dominant factor to reduce $\rho_c$ cannot be elucidated at present, it is definitely important to engineer the *DOS* of graphene to reduce $\rho_c$.

**Acknowledgements**

We thank Drs. Nabatame & Narushima, NIMS, for the fabrication of the SiN membrane masks. We are grateful to Covalent Materials for kindly providing us with the Kish graphite. This work was partly supported by the JSPS through its "Funding Program for World-Leading Innovative R&D on Science and Technology (FIRST Program)", by a Grant-in-Aid for Scientific Research from the Ministry of Education, Culture, Sports, Science and Technology, and by the Semiconductor Technology Academic Research Center (STARC).

supplementary material

# The density of states of graphene underneath a metal electrode and its correlation with the contact resistivity

*Ryota Ifuku, *Kosuke Nagashio, Tomonori Nishimura, and Akira Toriumi*

Department of Materials Engineering, The University of Tokyo, Tokyo, 113-8656, JAPAN

*Corresponding author: nagashio@material.t.u-tokyo.ac.jp

**Figure A1(a)** shows the typical optical image of graphene on ~ 3-nm $SiO_2/n^+$-Si with the RGB adjustment. Graphene can be detected after the contrast is optimized by adjusting RGB values of the image (graphene on the TV monitor is more clear). Graphene is brighter than $SiO_2$ because of the negative optical contrast value, as shown in **Fig. 1(b)**. As shown in **Figure A1(b)**, its contours are clearly observed in the dark-field mode. By combining these two methods, the monolayer graphene was detected and confirmed by the Raman spectroscopy, as shown in **Fig. A1(c)**. Although the D peak of graphene is blinded by the peaks from the Si substrate (1200 - 1500 $cm^{-1}$), no D band is generally observed in graphene transferred from Kish graphite.

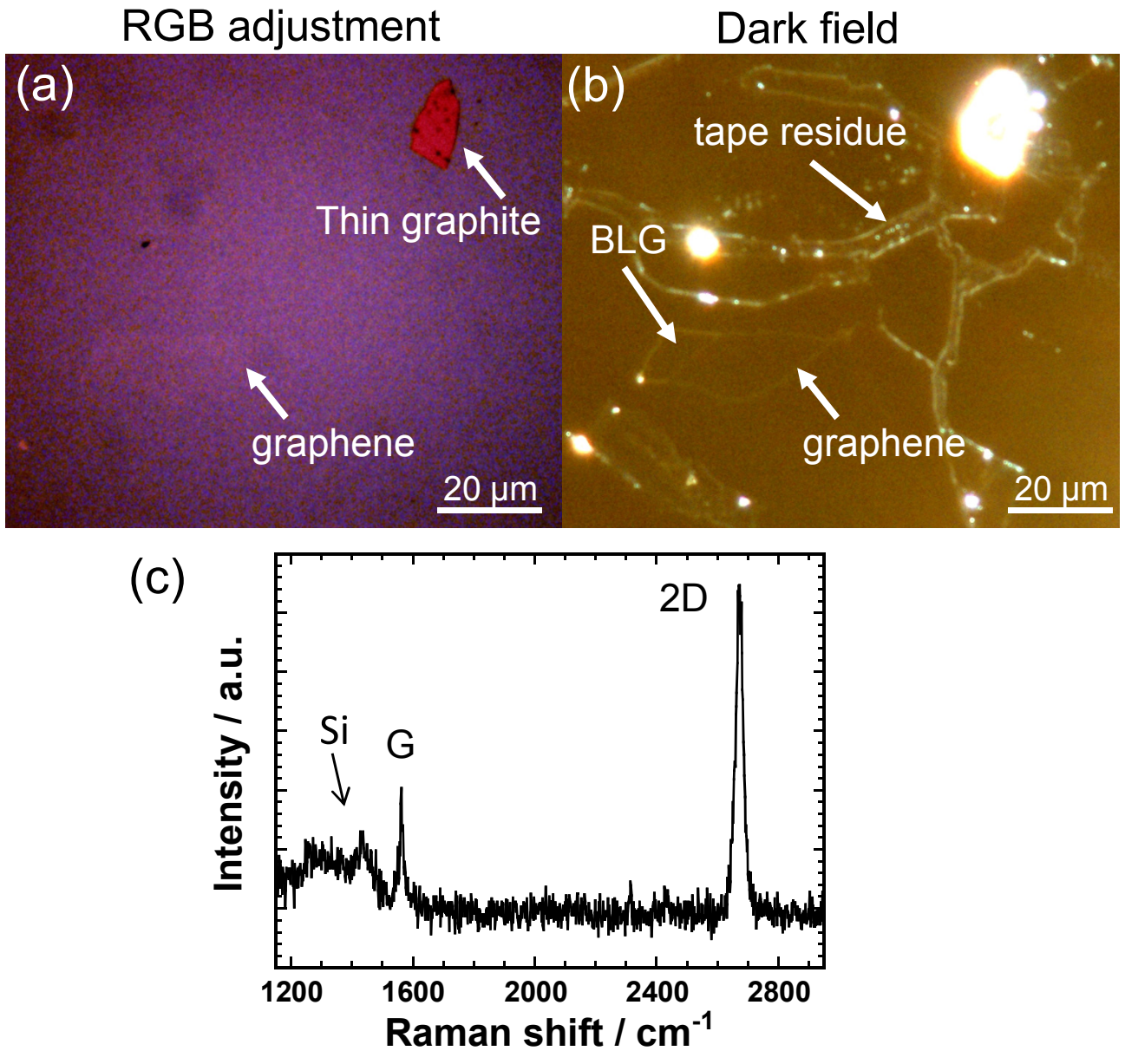

**Fig. A1 (a) Typical optical image of graphene on ~ 3-nm $SiO_2/n^+$-Si, (b) Dark field mode of (a). (c) Raman data of graphene on ~ 3-nm $SiO_2/n^+$-Si.**

For the resist-free metal deposition, the pre-patterned PMMA stencil mask supported by the Si substrate with the 200-μm square window was fabricated, as shown in **Fig. 2(b)**. **Figure A2** shows the fabrication sequence of this PMMA mask. (a) PVA and PMMA were spin-coated on the Si wafer with 1-cm square size. (b) The patterns required for *C-V* measurement were drawn by the electron beam li-thography. (c) The Si wafer with PMMA mask was soaked in water to dissolve PVA. (d) The floating PMMA was caught by the Si wafer with a 200-μm square window. Finally, it was dried on the oven.

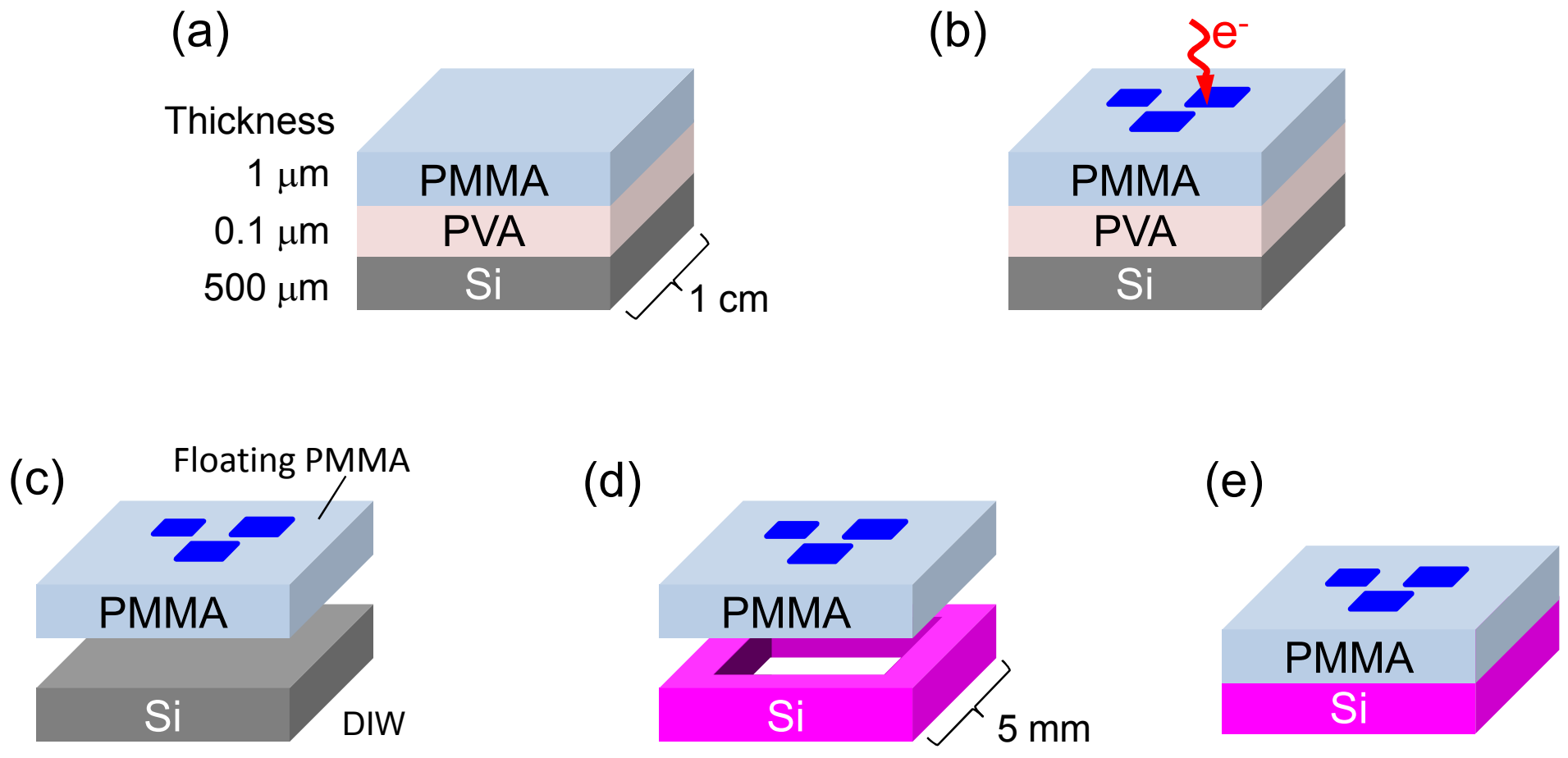

**Fig. A2 Fabrication sequence of the PMMA mask.**

The larger asymmetry of the resistance was observed especially for resist-free Ni device, as shown in **Fig. 4**. **Figure A3** shows the schematic of the band diagram showing (i) the charge transfer region, (ii) *p-n* junction & (iii) carrier density modulation just below Ni for (a) the resist-processed Ni device and (b) the resist-free Ni device. *T*he *p-n* junction formed near the contact always exists for $V_{BG} < 0$ V for resist-free Ni device, which results in the additional resistance. Therefore, the larger asymmetry of the resistance was observed for resist-free Ni device.

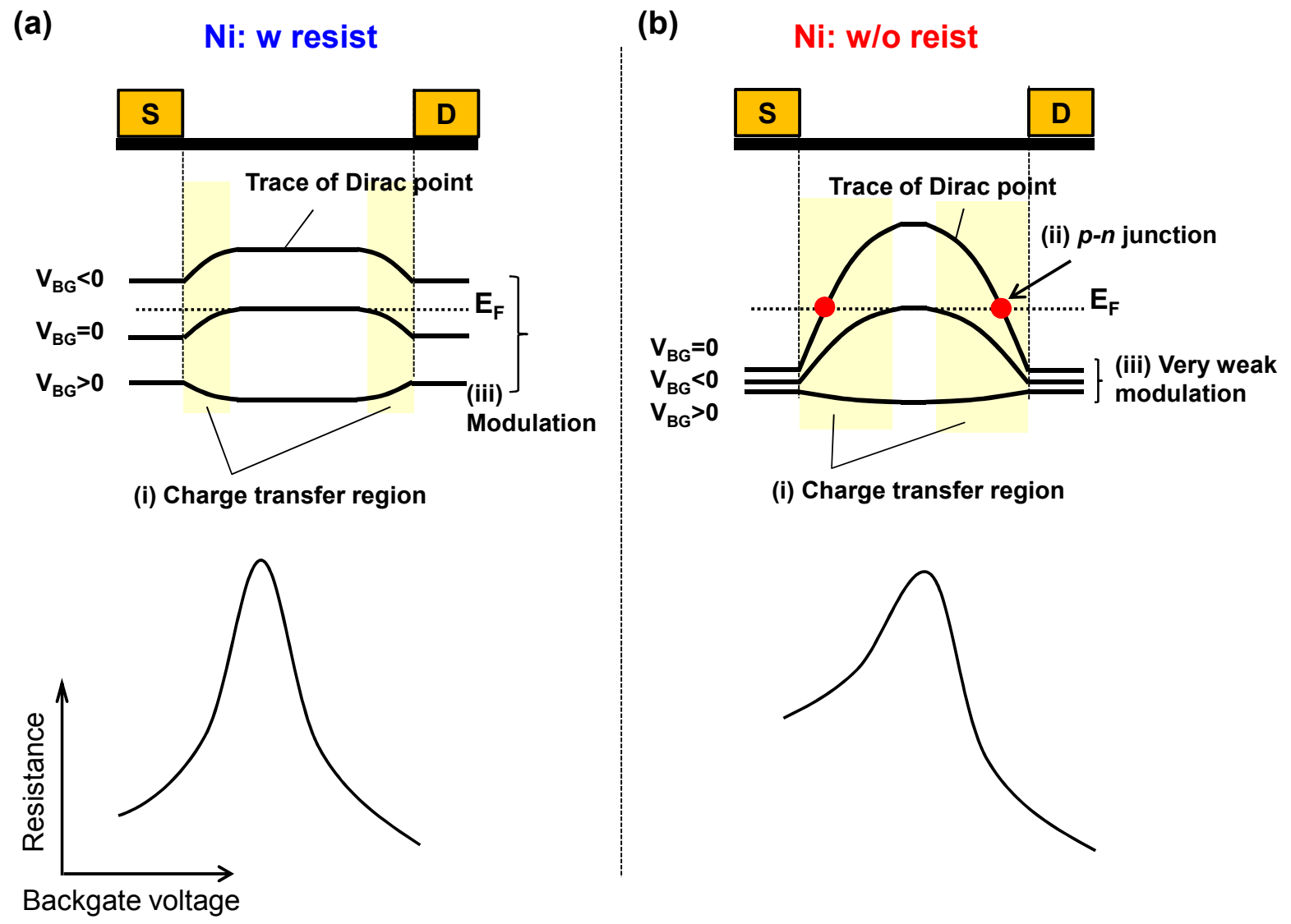

**Figure A3 Schematics of the band diagram and *I-V* curve for (a) the resist-processed Ni device and (b) the resist-free Ni device.**

**Figure A4** shows TLM analysis at different $V_{BG}$ for resist-free Au device. The fitted contact resistance ($2\times R_C$) was found to be ~20-60 Ω for different back gate biases. Because the width of the graphene channel is 1.4 μm, the unit length contact resistance is ~ 14-42 Ωμm for different back gate biases. It should be noted that the data for the Dirac point (~0 V) is neglected since the absolute errors around the Dirac point are significantly large due to the large resistance.

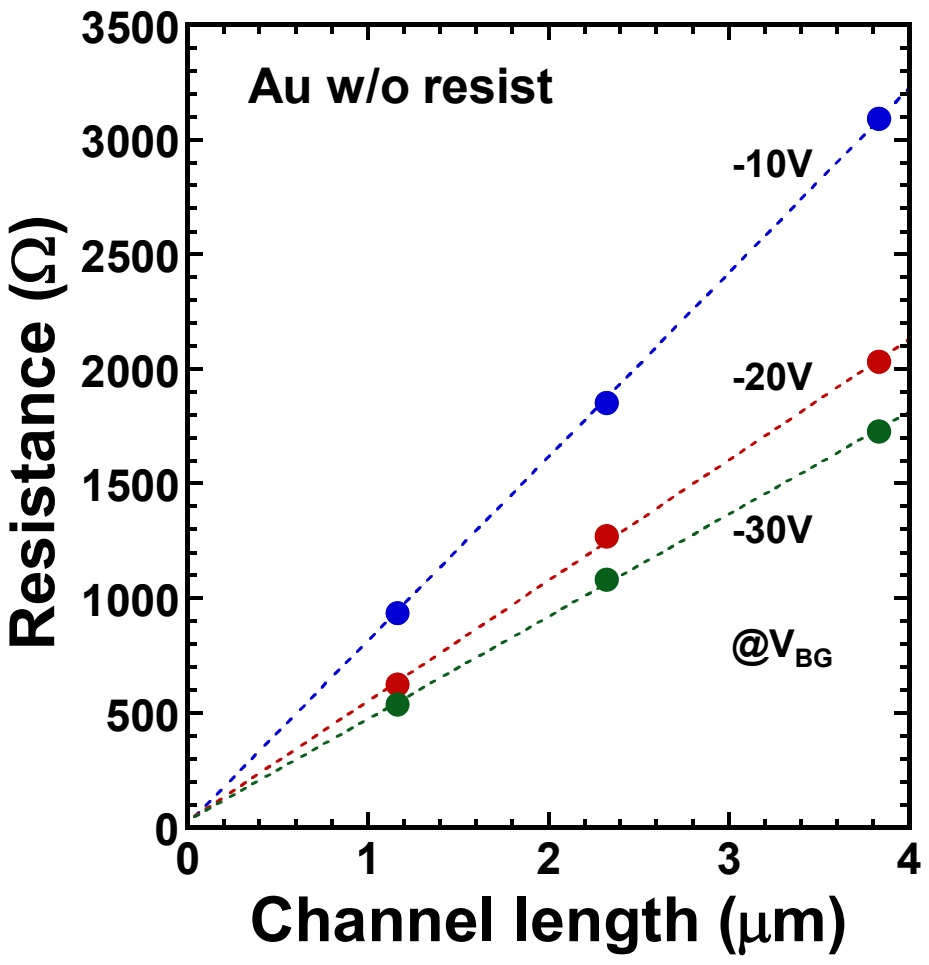

**Fig. A4 TLM analysis at different $V_{BG}$ for resist-free Au device.**